\documentclass{article}
\usepackage{arxiv}

\usepackage{graphicx}
\usepackage{dcolumn}
\usepackage{bm}

\usepackage{graphicx,amsmath,amsfonts,amscd,amsthm,amssymb,pstricks}
\usepackage{xcolor}
\usepackage{dcolumn}
\usepackage{bm}
\usepackage{mathrsfs}
\usepackage{textcomp}
\usepackage{esvect}
\usepackage{float}
\usepackage[USenglish,british]{babel}
\usepackage{environ}
\usepackage{xifthen}
\usepackage{xargs}		
\usepackage{hyperref}
\usepackage{physics,mathtools}
\usepackage[separate-uncertainty = true]{siunitx}
\usepackage{multirow}
\usepackage{comment}
\usepackage{enumerate}
\usepackage{tikz}	
\usetikzlibrary{arrows, calc, matrix, patterns, decorations.markings, shapes, decorations.pathmorphing, shadows.blur}
\usepackage[utf8]{inputenc}
\usepackage{enumitem}
\usepackage{caption,booktabs}
\usepackage{subcaption}
\title{Exploring the Potential of Resonance Islands and Bent Crystals for a Novel Slow Extraction from Circular Hadron Accelerators}

\author{D. E. Veres\\
CERN, Esplanade des Particules 1, 1121 Meyrin, Switzerland\\
Also at Goethe University, 60323 Frankfurt am Main, Germany\\
\And
G. Franchetti\\
GSI Helmholtzzentrum f\"ur Schwerionenforschung GmbH, Planckstraße 1, 64291 Darmstadt, Germany\\
Also at Goethe University, 60323 Frankfurt am Main, Germany\\
\And
M. Giovannozzi\thanks{Corresponding author: massimo.giovannozzi@cern.ch}\\
CERN, Esplanade des Particules 1, 1121 Meyrin, Switzerland}%

\begin{document}
\maketitle

\begin{abstract}
New developments in accelerator physics have broadened the set of available techniques for manipulating charged-particle beams. Adiabatic trapping and transport of beam in resonance islands has been studied and successfully implemented at the CERN Proton Synchrotron to perform multiturn extraction. Bent crystals have been successfully installed in the CERN Large Hadron Collider, improving the cleaning performance of the collimation system, and at the CERN Super Proton Synchrotron for reducing losses at the extraction septum in the case of slow extraction. In this paper, we discuss the potential of the combined use of resonance islands and bent crystals to devise a novel technique to perform slow extraction in circular \textcolor{black}{hadron} accelerators. The proposed approach is promising, particularly for applications with high-intensity beams, as it could dramatically reduce the losses on the extraction devices. 
\end{abstract}


Modern circular particle accelerators are based on a design rooted in seminal works that developed the concept of strong focusing~\cite{PhysRev.91.202.2,COURANT19581}. In this paradigm, the dynamics of charged particles is assumed to be linear, from which any departure necessitates correction techniques to mitigate nonlinear beam dynamics effects. Although considering nonlinearities to be harmful is partially correct, nonlinear beam dynamics can also open new possibilities for controlling and manipulating charged-particle beams. This is exemplified by Multi-Turn Extraction (MTE)~\cite{PhysRevLett.88.104801,PhysRevSTAB.7.024001}, developed at the CERN Proton Synchrotron (PS) and now the operational extraction mode of high-intensity fixed-target proton beams for the Super Proton Synchrotron (SPS)~\cite{Borburgh:2137954,PhysRevAccelBeams.20.061001,PhysRevAccelBeams.22.104002,PhysRevAccelBeams.25.050101}. The usefulness of exploiting nonlinearities is further illustrated by recently proposed advanced beam manipulations using transverse exciters~\cite{our_paper4,PhysRevAccelBeams.26.024001} or the crossing of 2D resonances~\cite{our_paper7,capoani:ipac23-wepl098}. \textcolor{black}{While proven successful, such manipulations have not yet been widely adopted in operation outside CERN and still present challenges that merit further studies.}

Another cutting-edge topic in accelerator physics is the use of bent crystals to control the trajectory of particles~\cite{lindhard:1965}. After extensive research~\cite{carrigan:1990,elsener:1996,biryukov:1997,PhysRevLett.87.094802,PhysRevSTAB.6.033502,fliller:2005,PhysRevLett.98.154801,PhysRevSTAB.9.013501,PhysRevSTAB.11.063501,PhysRevLett.102.084801,scandale:2010,scandale:2013,PhysRevAccelBeams.21.014702,shiltsev:2019,PhysRevApplied.14.064066,scandale:2022}, bent crystals have become an essential part of the operational collimation system of the CERN Large Hadron Collider (LHC)~\cite{PhysRevAccelBeams.27.011002}. Several concepts of crystal-assisted slow extraction have also been explored, such as non-resonant extraction of the beam halo~\cite{Fraser:IPAC2017-MOPIK048} or shadowing of the electrostatic septum using bent crystals~\cite{velotti:ipac17-mopik050,esposito:ipac19-wepmp028,velotti:ipac19-thxxplm2,PhysRevAccelBeams.22.093502,velotti:ipac22-wepost013}. Furthermore, the use of crystals to extract the beam halo to use for fixed-target experiments in the LHC~\cite{fomin:2017,bagli:2017,botella:2017,redaelli:ipac18-tupaf045,PhysRevLett.123.011801,fomin:2019,mirarchi:2020,Dewhurst:2023cth} is currently being investigated by the Physics Beyond Colliders initiative at CERN~\cite{Barschel:2653780,jaeckel:2020,pbc}. We present here a novel beam manipulation that combines stable islands, generated by sextupoles and octupoles, and bent crystals to realize an efficient alternative to the standard slow extraction technique.

Standard slow extraction uses third-order resonance driven by sextupoles to transport particles along unbounded separatrices to higher amplitudes in the $x$ plane. The thin blade of an electrostatic septum is used to cut the distribution along one separatrix arm and deflect some particles toward the extraction channel (see, e.g., Refs.~\cite{Tuck:1951,LeCouteur:1951,gordon:1958,hammer:1961,kobayashi:1967,gordon:1971,PIMMSvol1,Hardt:1025914,PulliaSX} and references therein). This process is characterized by the inherent loss of the beam on the septum that produces irradiation that can reduce the useful life of the accelerator components, hinder maintenance~\cite{Fraser2017JACoWS, PhysRevAccelBeams.22.123501}, and limit the total intensity of the beam that can be extracted. Additionally, such an extracted beam has a horizontal profile that is often difficult to match with the subsequent transfer line. The novel approach presented here has the potential to improve both aspects. 

\begin{figure*}
\includegraphics[width=\textwidth]{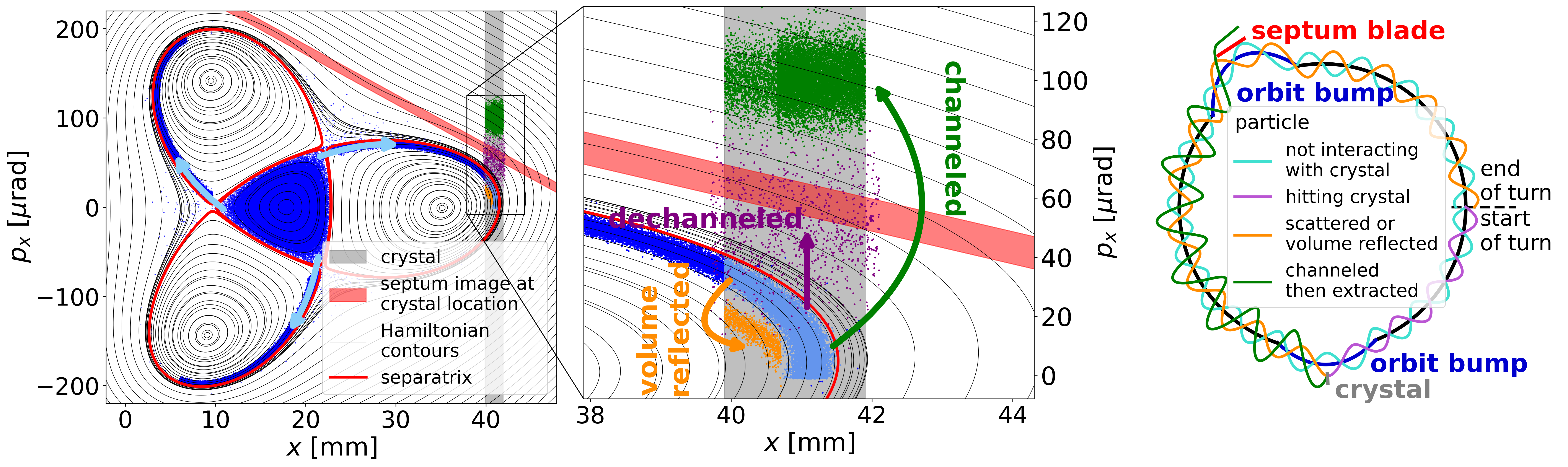}
\caption{\label{fig:principle} Schematic illustration of the new proposed slow extraction technique. Left: Example phase space at the crystal position, including the septum blade image. The core particles move into the resonance islands and follow the separatrix until they reach the crystal, where different interactions can occur (center). Right: Path of particles around the ring before and after various interactions with the crystal. A channeled particle (green) can jump the septum, while other particles (cyan and orange) continue to circulate.}
\end{figure*}

In the proposed approach, sextupole and octupole magnets generate stable islands in the horizontal phase space as the tune adiabatically approaches the third-order resonance. During the process, particles are captured in the islands in a controlled manner and move close to the separatrix until they reach a bent crystal, as illustrated in Fig.~\ref{fig:principle}. \textcolor{black}{Their motion is regular, unlike in the standard slow extraction, in which the particles have a stochastic dynamics}. Within the crystal, the particles can undergo different types of interactions (shown in Fig.~\ref{fig:principle}, discussed later). For the presented approach, the most relevant is planar channeling, which bends particles to a higher angle, allowing them to jump beyond the septum blade\footnote{When discussing hardware for particle accelerators, the term blade is customarily referred to the thick separation in a magnetic septum, while in the case of an electrostatic septum, a thin foil or a line of wires is used. In this paper, the term blade will be used with a generic meaning, which does not necessarily imply a magnetic device.}. Other interactions experienced by particles with less favorable initial conditions, such as volume reflection~\cite{PhysRevLett.97.144801} or slight scattering, do not significantly alter their subsequent path in the ring, so they will remain close to the separatrix. As a result, they have the possibility of multiple passes through the crystal.\textcolor{black}{ This is made possible by the use of stable islands and is in contrast to conventional slow extraction where the motion outside the core is unstable, making multi-pass channeling impossible. Ensuring multiple passes through the crystal is essential, as this} increases the crystal efficiency and facilitates low-loss beam extraction. The slow change of the system's parameters, allowing trapping in the islands, generates a variation in the size of the islands. Two variable-amplitude orbit bumps keep the horizontal beam position fixed at the crystal and the extraction septum. 

The principle was studied in numerical simulations using a one-turn map representing a simple model of nonlinear transverse motion in a circular accelerator lattice in the presence of sextupoles and octupoles that are assumed to be at the same location and are represented using the one-kick approximation~\cite{Bazzani:262179}. This map can be represented in 4D Courant-Snyder coordinates~\cite{COURANT19581} as
\begin{equation}
\begin{split}
    &\left(\begin{matrix}
    \hat{x} \\
    \hat{p}_x \\
    \hat{y} \\
    \hat{p}_y
    \end{matrix}\right)_{n+1}=
    R(2\pi Q_x,2\pi Q_y)
    \left(\begin{matrix}
    \hat{x} \\
    \hat{p}_x+f_x(\hat{x},\hat{y}) \\
    \hat{y} \\
    \hat{p}_y+f_y(\hat{x},\hat{y})
    \end{matrix}\right)_n \, , \\
    &f_x(\hat{x},\hat{y})=\frac{K_2}{2}\beta_x^{3/2}(\hat{x}^2-\chi\hat{y}^2)+\frac{K_3}{6}\beta_x^2(\hat{x}^3-3\chi\hat{x}\hat{y}^2) \, , \\
    &f_y(\hat{x},\hat{y})=-K_2\beta_x^{3/2}\chi\hat{x}\hat{y}+\frac{K_3}{6}\beta_x^2(\chi^2\hat{y}^3-3\chi\hat{x}^2\hat{y}) \, ,
\end{split}
\label{eq:map_4D}
\end{equation}
where $R$ represents the $4\times4$ rotation matrix, obtained as the direct product of two $2\times 2$ rotation matrices, namely $R(2\pi Q_x,2\pi Q_y)=R(2\pi Q_x)\otimes R(2\pi Q_y)$, with $Q_x$ and $Q_y$ being the horizontal and vertical betatron tunes, respectively. $K_n=\frac{1}{B_0\rho}\frac{\partial^nB_y}{\partial x^n}L$ are the integrated normal multipole strengths, where $L$ represents the length of the multipolar element, $B_y$ the vertical component of its magnetic field, and $B_0\rho$ the magnetic rigidity. The horizontal and vertical $\beta$-functions are represented by $\beta_x$ and $\beta_y$, respectively, with $\chi=\beta_y/\beta_x$. The parameters used in the numerical simulations were chosen based on the SPS~\cite{Adams:1970jz,Adams:1220684}. $K_2$ was set to \SI{0.0722}{\meter^{-2}}, while horizontal and vertical $\beta$-functions were set to \SI{104}{m} and \SI{20}{m}, respectively. Note that for $\chi \ll 1$ ($\chi=0.19$ in our case) the effect of the vertical plane can be neglected, since the two degrees of freedom are only weakly nonlinearly coupled if $Q_y$ is far from any low-order resonances. Nevertheless, for numerical simulations, the vertical degree of freedom was included, except for in the semi-analytical calculations originating from the Normal Form analysis (see below). The simulations used a transverse Gaussian beam distribution with normalized horizontal and vertical emittances $\epsilon^\ast_x = \SI{10}{\micro \meter}$ and $\epsilon^\ast_y = \SI{3.6}{\micro \meter}$, respectively, and a momentum $p_\mathrm{beam} = \SI{400}{GeV/\textit{c}}$ corresponding to that of the SPS fixed-target proton beam.

The implementation of these ideas requires precise control of the separatrices, and an analytic model of their dependencies is necessary, which can be provided by Normal Form theory. Using a 2D equivalent of map~\eqref{eq:map_4D}, describing the motion in the $x$ plane where all beam manipulations take place, the Normal Form technique was used according to the approach developed in Ref.~\cite{Bazzani:262179} to obtain a time-independent Hamiltonian. The phase flow of this Hamiltonian interpolates the particle orbits at integer times, and can be represented in the action angle coordinates $(\rho,\theta)$ as
\begin{equation}
\begin{split}
    \mathcal{H}&=\epsilon\rho+\frac{\Omega_2}{2}\rho^2-\frac{\epsilon \cos 3\theta}{4\sin\left(\frac{3\epsilon}{2}\right)}\rho^{3/2} \, , \\
    \frac{\epsilon}{2\pi} &= Q_x-\frac{1}{3}\, , \quad \Omega_2 =\Omega_2(Q_x,\kappa) \, , \quad \kappa=\frac{2}{3}\frac{K_3}{K_2^2}\frac{1}{\beta_x} \, ,
\end{split}
\label{eq:Hamiltonian}
\end{equation}
where $\Omega_2$ has a complicated dependence on $Q_x$ and the multipole strengths via $\kappa$. Analysis of the Hamiltonian~\eqref{eq:Hamiltonian} allows us to determine the equation of the separatrices as a function of $Q_x$ and $\kappa$.

A key element of the approach is the trapping of particles in stable islands. In a time-independent system, the separatrix represents an impenetrable boundary. Therefore, it is essential to introduce a time dependence of some system parameters to allow crossing of the separatrix and trapping in the islands. In the above model, $Q_x$ and $\kappa$ are obvious candidates for varying the shape, surface, and location of the different regions in the phase space. Trapping and transport processes in the adiabatic regime occur according to probabilistic rules that have been fully determined~\cite{neish1975,NEISHTADT198158,NEISHTADT1986,NEISHTADT1991,NEISHTADT2006158} for the Hamiltonian case and verified for the case of maps~\cite{PhysRevE.89.042915,our_paper4}. The necessary condition for non-zero trapping probability in a given phase-space region is a positive rate of change of its area. Another important consideration in constructing the appropriate variation of the system parameters is the rate at which particles cross the separatrix. To ensure a good extraction spill, it is desirable to decrease the intensity of the core linearly to zero during the process.

\begin{figure}
\centering
\includegraphics[width=0.8\linewidth]{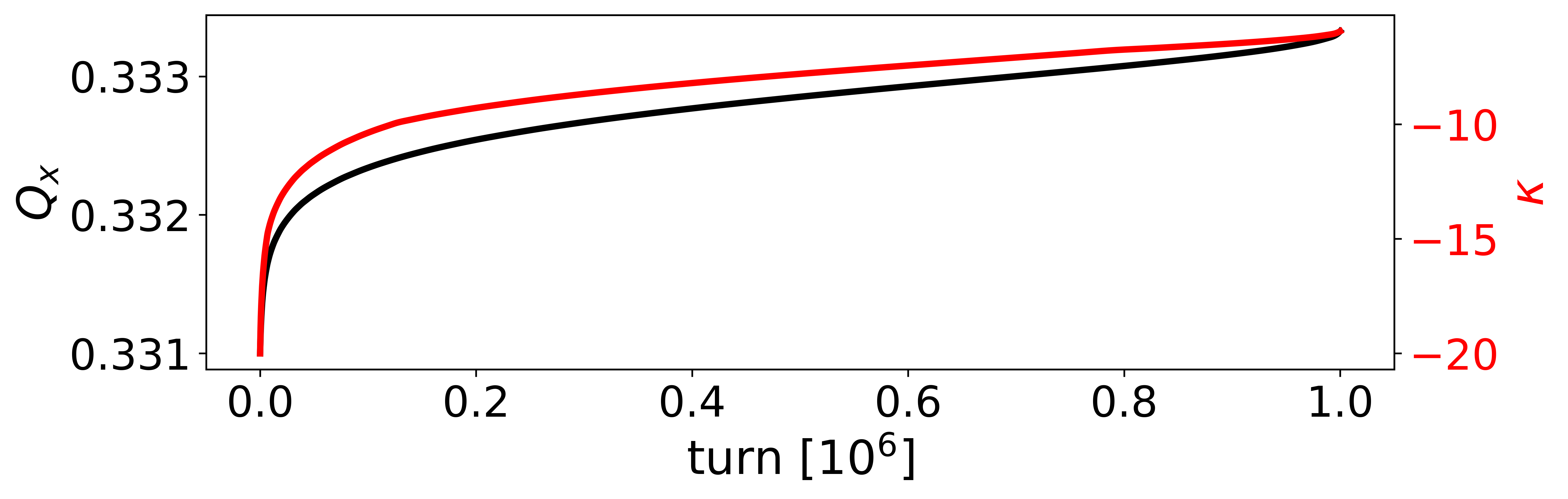}
\caption{\label{fig:tune_kappa_var} Variation of system parameters $Q_x$ (black) and $\kappa$ (red) as a function of the turn number used in the numerical simulations.}
\end{figure}

Taking into account these constraints, the variation of $Q_x$ and $\kappa$ shown in Fig.~\ref{fig:tune_kappa_var} was constructed, which provides the integrated octupole strength $K_3$ ($K_2$ is kept constant). The initial Gaussian distribution of the particles in the horizontal plane was tracked from far away from the resonance to the starting point of extraction using a linear change in tune and constant $\kappa$, to allow filamentation in the triangular-shaped core region. The Hamiltonian was then used to establish the relationship between $\kappa$ and $Q_x$ that ensures faster growth of the island surface than a decrease in the core surface as the tune approaches the resonance from below. Using the Hamiltonian again, the separatrix was calculated as $\kappa$ and $Q_x$ were varied to obtain the desired number of particles within the core as the resonance approaches, assuming the adiabaticity of the process. The most appropriate variation of $Q_x$ (and thus $\kappa$) was then constructed by prescribing a linear decrease in the intensity of the core. Note that, depending on the direction of approach of the resonance (from below or above), it is also possible to achieve the constraints imposed by varying only $Q_x$. All this allows particle transport to large amplitudes, preserving the action of their orbit.

The second key element of the proposed scheme is the use of a bent crystal to create the angular separation between the particles circulating and those to be extracted. This is achieved through planar channeling, however, particles can experience a variety of other interactions with the crystal depending on their entry conditions. An overview of these is given below (for a more detailed description of physics, see~\cite{Biryukov,SCANDALE20191}).

\begin{figure}
\centering
\includegraphics[width=0.8\linewidth]{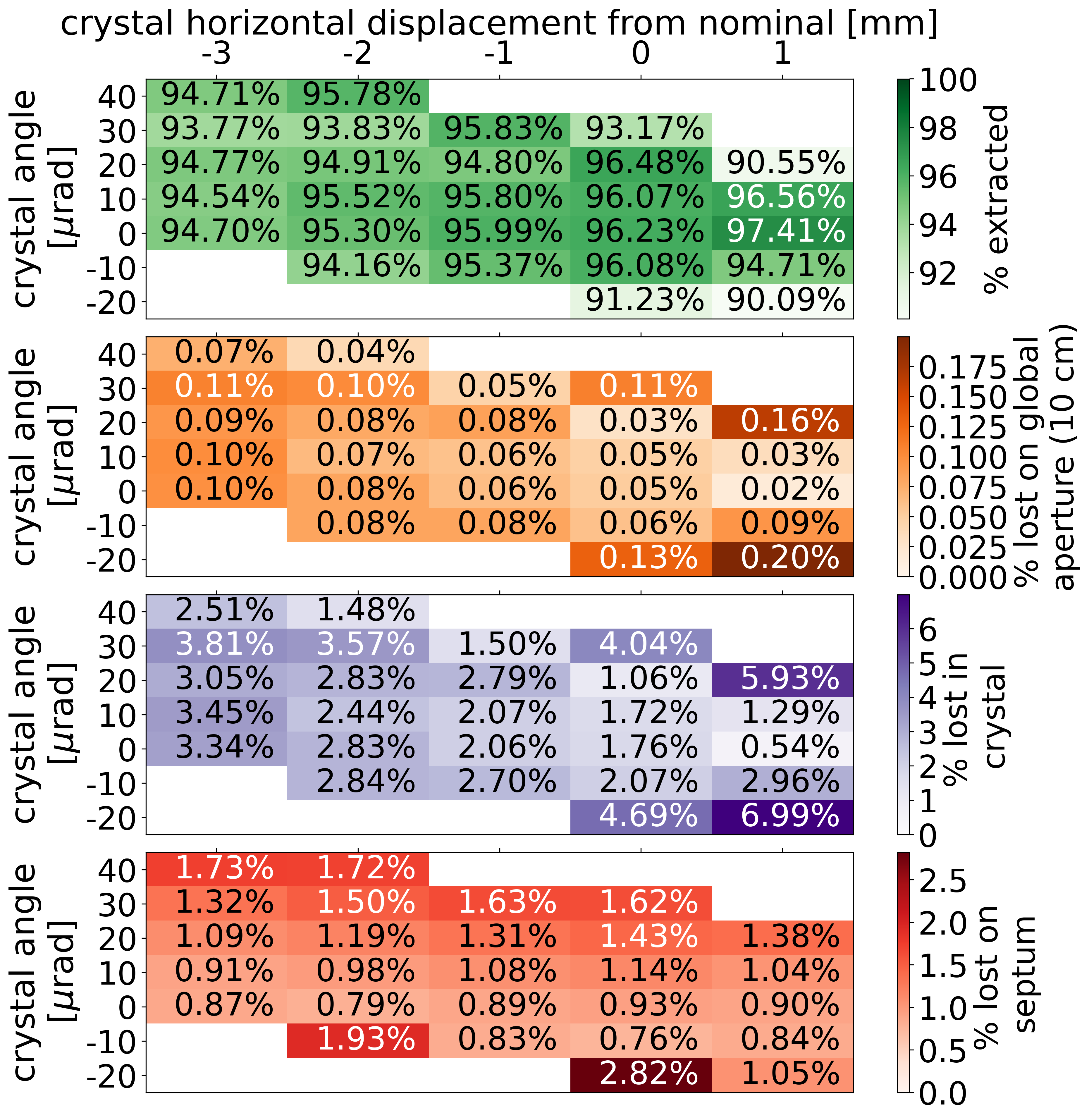}
\caption{\label{fig:scan_eff} From top to bottom: extraction efficiency, and losses on the mechanical aperture, in the crystal, and in the septum as a function of the angular and position alignment of the crystal with respect to the island separatrix. The remaining particles continue to circulate at the end of the process and can be extracted by pushing the horizontal tune closer to resonance.}
\end{figure}

Inside the crystal, the electric fields of the atomic planes form potential wells in which an incident charged particle can be trapped: this is called channeling. In a bent crystal with a macroscopic radius of curvature $R$, channeled particles oscillate between atomic planes following the curvature and emerge with an angle close to the crystal bending angle. Channeling occurs only if the particle’s transverse momentum remains below the maximum of the potential well. This implies a maximum impact angle, called the critical angle $\theta_\mathrm{c}$, which depends only on the crystal potential, $U_\mathrm{max}$, the particle momentum, $p$, and speed, $v$, and is given by $\vert\theta_\mathrm{c} \vert=\sqrt{\frac{2U_{\text{max}}}{pv}}\left(1-\frac{R_\mathrm{c}}{R}\right)$~\cite{Biryukov}, where $R_\mathrm{c}$ is the critical radius below which channeling is impossible.

During its motion in the crystal, a particle can scatter off electrons, nuclei, or crystal defects, causing its transverse energy to change. Thus, channeled particles may lose channeling conditions if their transverse energy increases above the potential well maximum because of scattering. This is called dechanneling. The reverse process, called volume capture, is also possible. A particle can also be reflected from the crystal plane when it impinges with a tangential momentum (volume reflection). Finally, the particle may interact entirely amorphously with the crystal~\cite{Mirarchi:2694336}.

The numerical simulations presented here used the crystal implementation within the {\tt Xcoll} module of the {\tt Xsuite} simulation package~\cite{iadarola2023xsuite,Xsuite}. The implementation is an exact transplantation of the crystal routine in {\tt SixTrack}~\cite{Mirarchi:2694336, MIRARCHI2015378, SixTrack}, which was compared with the results of the beam tests carried out in the CERN North Area (NA) and measurements in the LHC~\cite{Mirarchi:2036210,MIRARCHI2015378,Mirarchi:IPAC2014-MOPRI111,Rossi:2644175,PhysRevAccelBeams.27.011002,CAI2024169038}. The crystal parameters used in the simulations were chosen within a range proven effective by NA tests and the LHC collimation system~\cite{LHCcollweb}: horizontal and vertical dimensions of \SI{2}{\milli \meter} and \SI{5}{\centi \meter}, respectively, length of \SI{2.8}{\milli \meter}, and $R=\SI{28}{\meter}$, resulting in a bending angle of \SI{100}{\micro \radian}. With these parameters, a simulated single-pass channeling efficiency of around \SI{68}{\%} can be achieved for particles with incident angle $\vert p_x \vert \lesssim \vert \theta_\mathrm{c}\vert$, where $\vert \theta_\mathrm{c} \vert \approx \SI{10}{\micro\radian}$ for $p_\mathrm{beam}=\SI{400}{GeV/\textit{c}}$.

\begin{figure}
\includegraphics[width=\linewidth]{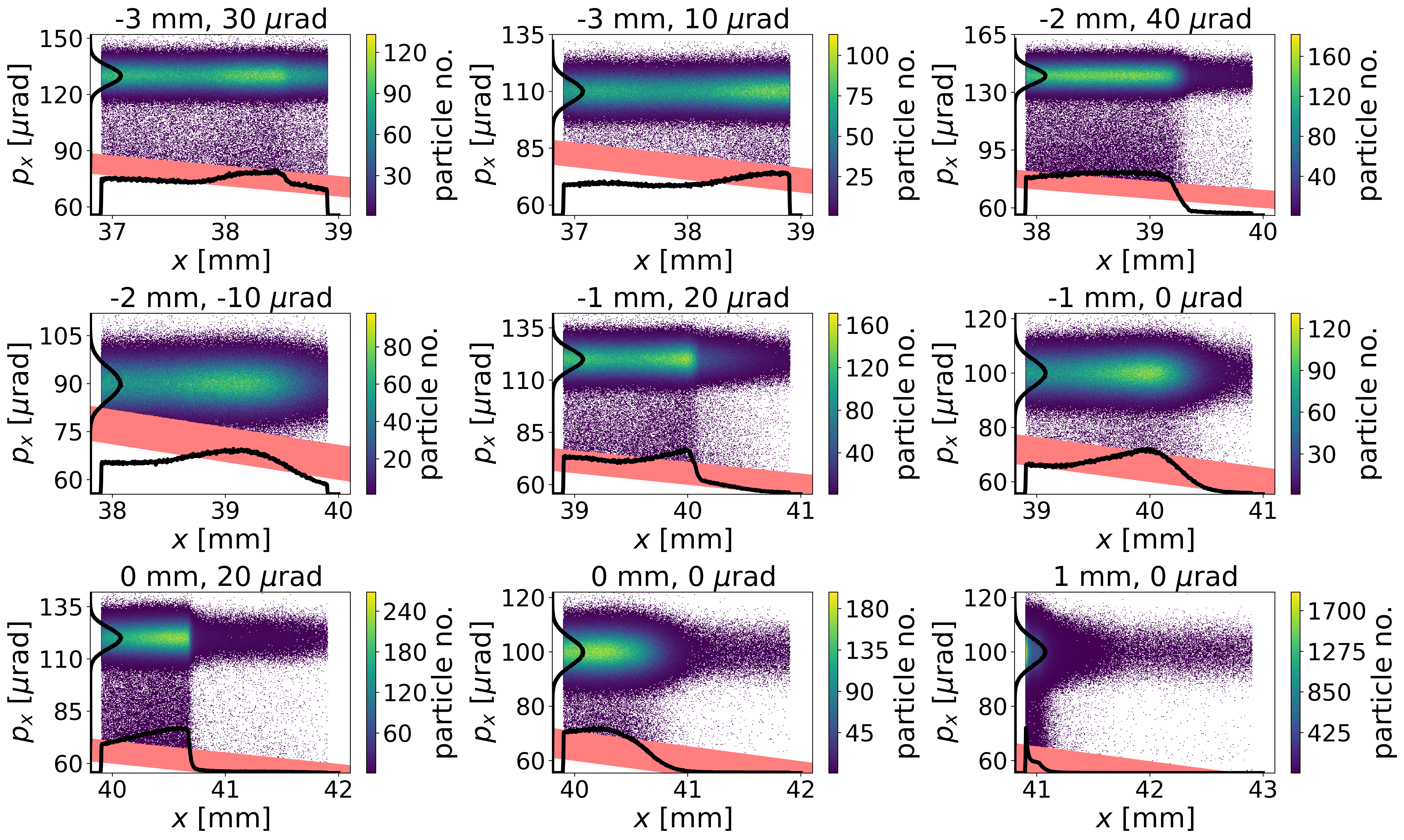}
\caption{\label{fig:scan_dist} Horizontal phase-space distribution (color) of the extracted particles at the crystal position for various angular and position alignments of the crystal with respect to the island separatrix. Projections along $x$ and $p_x$ are also shown (black), as well as the image of the septum blade (red).}
\end{figure}

Figures~\ref{fig:scan_eff} and \ref{fig:scan_dist} show the efficiency and shape of the extracted beam distribution at the crystal as a function of the crystal's angular and position alignment with respect to the island separatrix. The nominal crystal position is at \SI{40}{\milli\meter}, so that the extremum of the separatrix falls roughly in the middle of the crystal width, and the nominal angular alignment of the crystal is horizontal. Particles were considered lost in the septum if they hit the phase-space region covered by the image of the septum blade, backtracked by an optimal \SI{60}{\degree} phase advance through linear SPS optics, to the crystal location. The apparent width of the septum was taken as \SI{500}{\micro \meter}, consistent with the observed width of the electrostatic septum of SPS~\cite{PhysRevAccelBeams.22.093502, PhysRevAccelBeams.23.023501}. Particles that exceeded a radius of \SI{10}{\centi\meter} in the $x-y$ plane were considered lost in the mechanical aperture of the ring.

The efficiencies achieved in these simulations are comparable to or, in some cases, slightly exceed the measured slow extraction efficiencies in the SPS in 2016-2017~\cite{Fraser:IPAC2018-TUPAF054}. With careful crystal alignment, it is possible to reduce septum losses below 1\%. To minimize these losses, the crystal should be aligned so that the distribution of channeled particles is as close to the septum blade as possible without intersecting it. Increasing the separation between the channeled distribution and the septum slightly increases the septum losses as a higher density of dechanneled particles hits the blade. \textcolor{black}{The biggest novelty of the proposed extraction scheme is that it }can operate even without a septum, provided that the crystal can be placed in the lattice so that the channeled particles arrive at the right position and angle to the extraction channel. In this case, dedicated collimation of the dechanneled particles may be necessary. \textcolor{black}{The possibility of septum-less slow extraction is particularly exciting due to significant operational improvements it would present.} It is worth mentioning that this study assumed a simple accelerator lattice composed of periodic FODO cells. It is reasonable to expect further performance improvements if dedicated insertions with optimized optical configurations are implemented.

As Fig.~\ref{fig:scan_dist} illustrates, the profile of the extracted distribution can vary depending on the crystal alignment. To some extent, it is possible to tune the profile without significantly degrading efficiency, thus generating an extracted beam distribution that can be better matched to a transfer line or experimental requirements. In all cases, a low-density halo of particles scattered by the crystal is formed at higher angles or amplitudes that extend beyond the septum blade. In the most efficient configuration, this comprises less than 0.2\% of extracted particles. In reality, these particles would most likely be lost on the mechanical aperture of the ring or the transfer line, slightly affecting the overall efficiency. However, a dedicated collimation of these particles can be easily envisioned in the ring, mitigating any risk to the accelerator.

\begin{figure}
\centering
\includegraphics[width=0.8\linewidth]{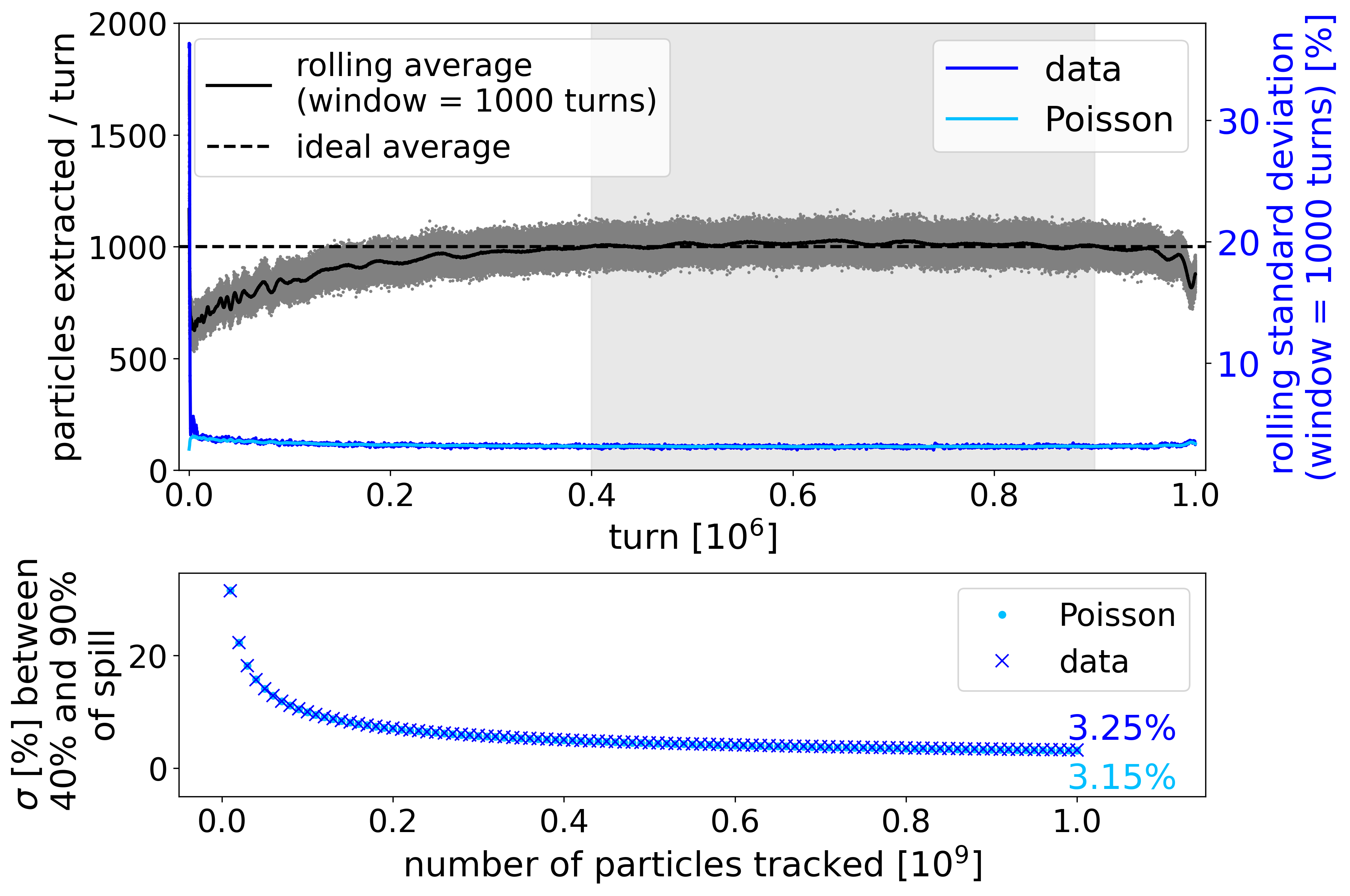}
\caption{\label{fig:spill} Top: Simulated extraction spill with $10^9$ particles and standard deviation of the number of particles extracted per turn. The most stable part of the spill is shaded in gray. Bottom: Standard deviation of the extracted spill in the shaded interval as a function of number of particles tracked.}
\end{figure}

Another important aspect of slow extraction is the temporal structure of the extraction spill. Fixed-target experiments require minimal variations in spill rate to avoid detector pile-up and other saturation phenomena. Figure~\ref{fig:spill} (top) shows a complete extraction spill, simulated using the nominal crystal alignment. Apart from the start of the spill, the average extracted particle count is constant within 2\% with a standard deviation of 3.25\%, closely approaching the Poisson limit. The lower quality of the initial part of the spill is due to the difficulty in steadily extracting the filamented tails of the transverse Gaussian distribution. Furthermore, the highest rate of change in system parameters occurs in this segment, resulting in insufficient adiabaticity and lower spill quality. In Fig.~\ref{fig:spill} (bottom), the standard deviation as a function of particle number used in the simulation in the most stable part of the spill, between 40\% and 90\% of the spill time (shaded area), follows the Poisson scaling. This demonstrates that the observed fluctuations are not related to the method, but rather to the simulation statistics. Note also that the level of fluctuations is comparable to that of standard slow extraction using Eq.~\ref{eq:map_4D} without the octupole and crystal.

In a real machine, power supply ripples cause tune modulation, strongly affecting the trapping, and thus the extraction efficiency and spill quality. Introducing a tune modulation similar to that observed in SPS~\cite{PhysRevE.55.3507}, with dominant frequencies of \SI{50}{\hertz} and its higher harmonics (see Ref.~\cite{PhysRevE.57.3432}) showed that only the lowest harmonics (\SI{50}{\hertz}, \SI{100}{\hertz}, \SI{150}{\hertz}) are harmful to the quality of the spill. This is also observed in simulations of standard slow extraction using Eq.~\ref{eq:map_4D} without octupole and crystal. Therefore, standard mitigation measures are necessary for these effects (see, e.g., \cite{PhysRevAccelBeams.22.072802}).

In conclusion, a novel slow extraction that combines stable islands and bent crystal is proposed. Simulations using a simple accelerator lattice model demonstrate its potential to reproduce or even exceed the efficiencies of standard techniques, while introducing the prospect of septumless slow extraction and the ability to shape the extracted distribution to match a subsequent transfer line. In a real accelerator, the use of multiple nonlinear elements with suitable phase advances would allow full control of the shape and location of the island, allowing the creation of narrower islands in which most particles hit the crystal within its angular acceptance, \textcolor{black}{likely} further improving efficiency.

It was also shown that the approach does not inherently degrade the quality of the spill compared to standard methods. However, active compensation of low-frequency power supply ripples will still be necessary to maintain good efficiency and spill quality.

Studies using a more realistic accelerator lattice are underway to mitigate any adverse effects of momentum-dependent tune offset and entry conditions at the crystal due to chromaticity and dispersion in the presence of a momentum distribution in the beam. Experimental tests are also being considered in the SPS.

\section*{Acknowledgements}
The authors express their gratitude to current and former colleagues from the CERN BE-ABP-NDC section, in particular Pascal Hermes and Stefano Redaelli, for their helpful insights on crystal physics, and to Frederik van der Veken for support with crystal simulations. We also thank Rebecca Taylor, Pablo Arrutia, Matthew Fraser, Francesco Maria Velotti, and members of the CERN SY-ABT-BTP section for the many useful discussions on the topic of slow extraction.
%
\bibliographystyle{unsrt}
\bibliography{mybib}
\end{document}